# Urban vegetation change after a hundred years in a tropical city (San José de Costa Rica)

Julián Monge-Nájera & Gabriela Pérez-Gómez

Laboratorio de Ecología Urbana, Vicerrectoría de Investigación, Universidad Estatal a Distancia, 474-2050 San Pedro de Montes de Oca, San José, Costa Rica; julianmonge@gmail.com, gabytta1985@gmail.com

**Abstract:** Urban vegetation is of key importance because a large proportion of the human population lives in cities. Nevertheless, urban vegetation is understudied outside central Europe and particularly, little is known about the flora of tropical Asian, African and Latin American cities. We present an estimate of how the vegetation has changed in the city of San José, Costa Rica, after about one century, with the repeat photography technique (based on a collection of 19$^{th}$ and early 20$^{th}$ century photographs by José Fidel Tristán and others) and with data from the Costa Rican National Herbarium. We found little vegetation change in the landscape of San José during the 20th century, where a total of 95 families and 458 species were collected in the late 19th and early 20th century. The families with most species were Asteraceae, Fabaceae, Poaceae, Lamiaceae, Euphorbiaceae, Solanaceae, Cyperaceae, Acanthaceae, Malvaceae, Piperaceae and Verbenaceae. Similar results have been found in Europe, where the number of plant species often is stable for long periods even when the individual species vary. Rev. Biol. Trop. 58 (4): 1367-1386. Epub 2010 December 01.

**Key words:** Urban flora, effects of urbanization, city landscape, photographic comparison, species list.

Urban vegetation has mostly been studied in central Europe, where about 50% of species are alien, half of them introduced before the 15$^{th}$ century. Despite the heavy traffic among European cities, five centuries have not been enough to homogenize their flora: the communities of species introduced after the year 1500 are characteristic of each city (Frank *et al*. 2008).

For animals, which particular species occur in cities is predicted by the "environmental filtering model" that in turn is based on plants. The model states that (1) there is natural selection of species living in urban ecosystems, (2) plants define key habitat characteristics and (3) habitats define which animals can live in the city (Williams *et al*. 2009). Generally, moderate urbanization produces some increase in plant biodiversity but is deleterious for invertebrates and mammals. A high level of urbanization is correlated with fewer species of plants, invertebrates, amphibians, reptiles, birds and mammals, possibly because humans willingly introduce plant species, but not animals, to their gardens (McKinney 2008).

Outside central Europe, urban vegetation is understudied but there are some recent data from Plymouth, England, where alien species increase with urbanization (Kent *et al*. 2001). Also in England, gardens in Sheffield have a total of 1 166 plant species (70% alien) and twice the garden size means 25% more species. In these gardens there are 63% biennials/perennials, 18% shrubs, 10% annuals and 8% trees (Smith *et al*. 2006).

In Anglosaxon and French North America, there is a surprising scarcity of recent studies on urban biotas, but some work has been done.

In the city of Halifax (Nova Scotia, Canada), soil moisture and light determine which species are present. Taxa adapted naturally to rock, grassland and flooded habitat find an analog habitat in the city and thrive (Lundholm & Marlin 2006).

The New York metropolitan region has 556 woody species and non-native invasive species are becoming more common (Clemants & Moore 2004). In the Pelham Bay Park, New York City, native species went from 72% to 60% and 26% of natives disappeared in 50 years (especially herbaceous and meadow-type plants, DeCandido 2004).

Even though recent studies are scarce, a meta-analysis found 79 studies of species richness with geographic data for New York City; of these, 17 studies found a decrease in species richness, six an increase and three found no change. However, all studies reported an increasing number of exotic species (Puth & Burns 2008).

Tropical cities are in areas of high biodiversity but little is known about Asian and especially African cities regarding urban flora. In Latin America the situation is better but worldwide no floral lists exist for the 50 most populated cities (Clemants 2002). In Jinan City, China, a methodological comparison found that gradient analysis from the urban center to the fringe gives better estimates of the urban flora than the traditional block-area analysis (Kong & Nakagoshi 2005). In Taipei, green areas have 164 tree species (few shared among sites) and large evergreen native species dominate. Larger parks have higher richness, more landscape fidelity to the original vegetation, and more rare and endemic species (Jim & Chen 2007). In Africa, the urban areas of the Nile Delta (Egypt) have vegetation that is mainly correlated with moisture, pH, fertility and texture gradients, but plants always occupy sites similar to their natural habitats (Shaltout & El-Sheikh 2002).

Latin America has a long history of scientific study of urban biota, particularly the plants and there are several recent studies from México, Peru, Brazil, Chile and Argentina.

In the city of Ensenada, Mexico, there are 161 species, 61% non-native (Garcillán *et al*. 2009). In Mexico City, trees are stressed from dry wind and unfavorable water flow caused by the pavement (Barradas 2000).

Brazil has the largest urban forest in world (Tijuca: 3 300 hectares) but it is being stressed by roads because roads are surrounded by invasive species that burn easily. The fires in turn open adjacent areas to more invasive vegetation and the damage spreads (Matos *et al*. 2002). There are very few studies of plants that grow on walls but in Jundiai, Brazilian, walls have a biodiversity of 28 species (dos Reis *et al*. 2006).

Normally, satellites are not used to study urban vegetation but in Arequipa, Peru, satellite images show that desert vegetation is being lost because of urban expansion (Polk *et al*. 2005). However, only ground work can reliably identify species and this kind of work has shown that temperate South America is not different from North America and Europe: at least half of the plant species in the Argentinean cities of Mendoza and Rosario are introduced. In Luján de Cuyo, Mendoza, 61 species were identified: 69% introduced (Méndez 2005). The vacant lots of Rosario each have one dominant species, a few abundant species and many rare species. Therophytes predominate and the proportions of indigenous and introduced species are similar (Franceschi 1996).

Chile is the Latin American country with the largest number of recent studies. Synanthropic communities in an urban footpath of Valdivia represents six associations and two communities (Finot & Ramírez 1998). In Concepción, green areas are dominated by non-native ornamental species (Paucharda *et al*. 2006). The distribution of urban vegetation reflects social inequalities. In Santiago, poor areas can have ten times less plant cover than rich neighbourhoods (Hernández 2008), similar to other countries (Pedlowski *et al*. 2002). However, workshops in poor areas of cities can result in an improvement of their vegetation (Garzón *et al*. 2004).

In Costa Rica, there is a long history of study of urban plants that began with the National Museum's collection efforts in the late 19th century, but little has been published. Méndez & Fournier (1980) and Monge-Nájera *et al*. (2002a,b) studied the lichens and their relationship with air pollution. The use of European lichens proved succesful when they were transplanted to this Tropical city (Grüninger & Monge-Nájera 1988). Francisco Fallas made checklists and abundance estimates of urban herbs in the late 1970's but to our knowledge he did not publish them. The program "Costa Rica: Jardín Botánico de América Tropical" produces manuals and labels for urban vegetation (www.hjimenez.org) and there is a program to provide urban parks with butterflies and their host plants (http://www.lrsarts.com/plas/index.html).

The biodiversity in patches of urban vegetation can be surprisingly high, at least in Costa Rica. For example, after 50 years, in only one hectare of urban vegetation in San José, there are 432 plant species (Di Stéfano *et al*. 1995, Nishida *et al*. 2009), a full new lichen family with a novel symbiotic lifestyle (*Eremithallus costaricensis*; Lücking *et al*. 2008), 200 butterfly species (Nishida *et al*. 2009) and a new species of Hymenoptera, *Meteorus oviedo* (Shaw & Nishida 2005).

We present an estimate of how the vegetation has changed in the city of San José, Costa Rica, after about one century, with a technique called repeat photography. We do not know of repeat photography studies on urban Costa Rican vegetation of San José, but the technique was used in Costa Rica by Horn (1989) to assess changes in the páramo habitat.

## MATERIALS AND METHODS

We used a collection of photographs taken in the late 19th and early 20th centuries from the José Fidel Tristán Fernández Collection in the Archivo Nacional de Costa Rica and others reproduced by Leiva (2004). The sites were re-photographed on December 11, 2008 with a Nikon Coolpix 8800 camera (8 megapixels; Fig. 1). The repeat photography technique, developed in 1880 (Webb *et al*. 2010) is good for detailed analysis (Hendrick & Copenheaver 2009) and is cost- effective (Robert *et al*. 2010). We used digital repeat photography, which is fast, detailed and reliable; can be stored for future corroboration and comparison; includes rich data that may become useful in the future; and can classify and measure information automatically (Crimmins & Crimmins 2008).

We re-photographed nine sites (exact year of original photograph included when known): **Site 1** Catedral Metropolitana Street: 0 Avenues: 2-4 (1896); **Site 2** Catedral Metropolitana St. 0 Av. 0-2 (*c*. 1910); **Site 3** Kiosco Parque Morazán St. 5-7 Av. 3 (1914); **Site 4** Colegio de Señoritas St. 3-5 Av. 4-6 (1914); **Site 5** Paseo Colón St. 36 Av. 0, looking east (1899); **Site 6** Paseo Colón St. 36 Av. 0 looking west; **Site 7** Antigua Casa Presidencial St. 15 Av. 7; **Site 8** Estación del Ferrocarril al Atlántico St. 21-23 Av. 3; **Site 9** Hospital de Niños St. 20 Av. 0.

We calculated % cover by clipping and weighing sections from photographs printed on standard bond paper; for example, if the clippings from buildings represented 20 % of the total weight of the photograph, we recorded that buildings represented 20% of the image. This technique is compared with others by Monge Nájera *et al*. (2002b).

We also present an analysis of plants collected in cantón de San José from 1885 through 1945 by the Museo Nacional staff (Base de Datos, Herbario Nacional, updated to April 13, 2009).

## RESULTS

After about a century, the main change in San José city photographs is the much larger number of people. The reduction in vegetation affects grasses, shrubs and trees, but is small; the increase in buildings, streets, vehicles, sidewalks, lamps and signs is also small (Fig. 2).

A total of 95 families and 458 species were collected in the late 19th and early 20th century.

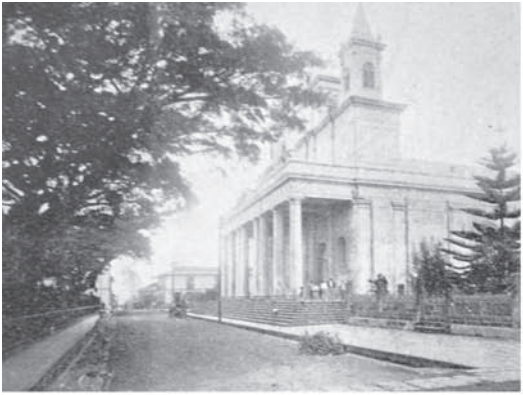
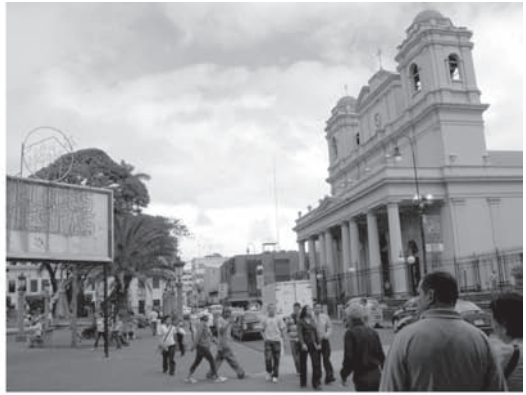
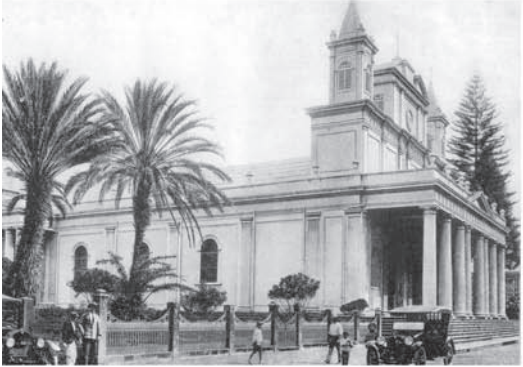
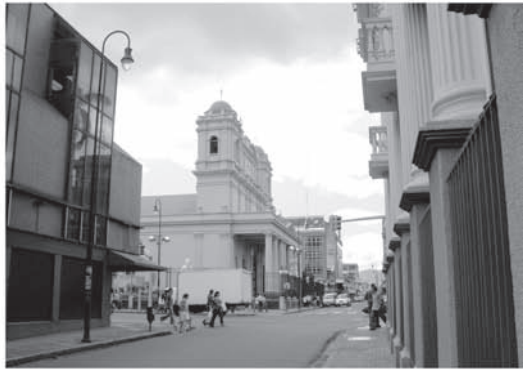
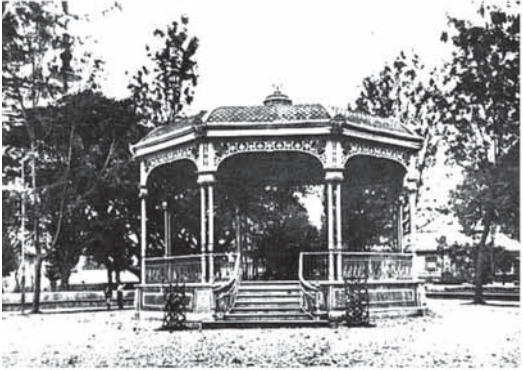
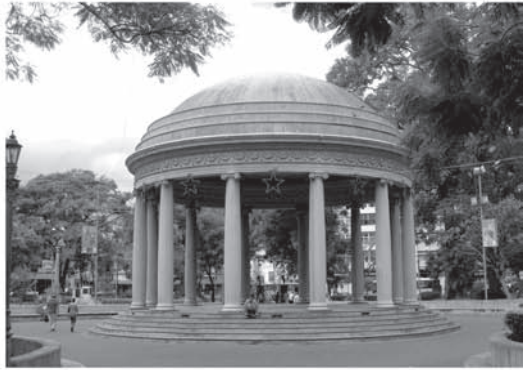
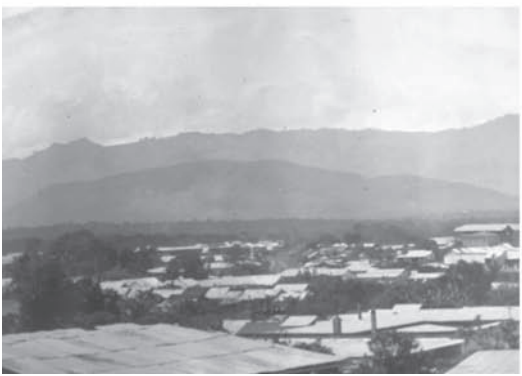
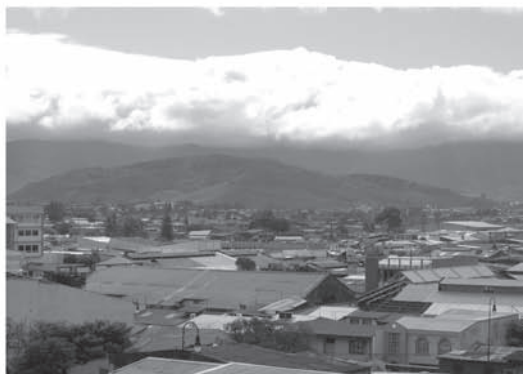

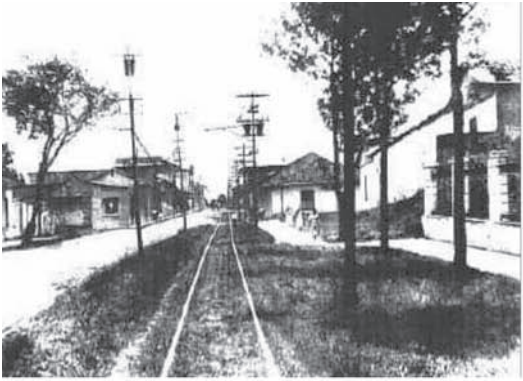 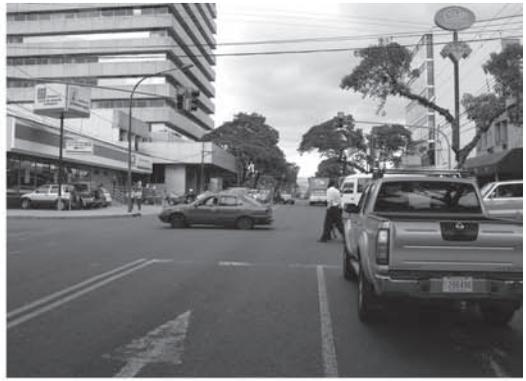

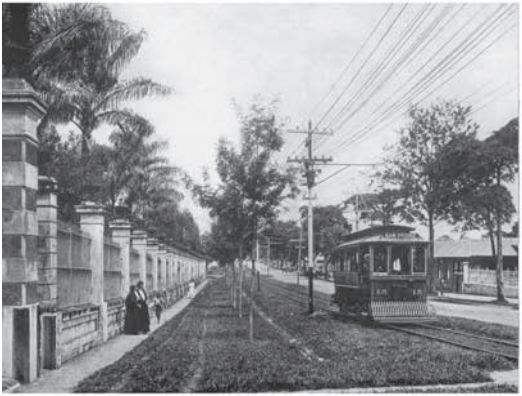 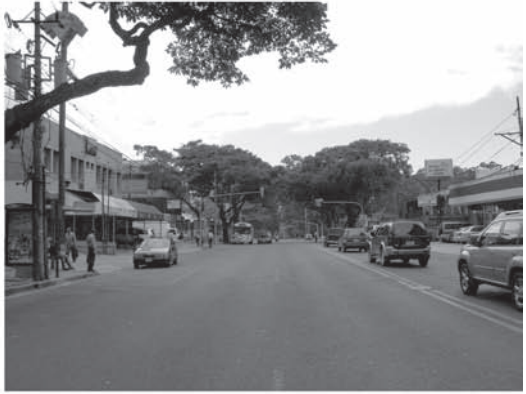

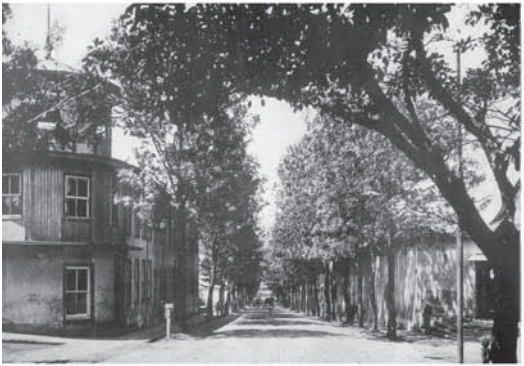 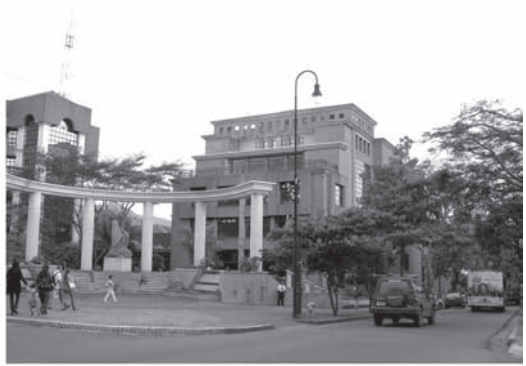

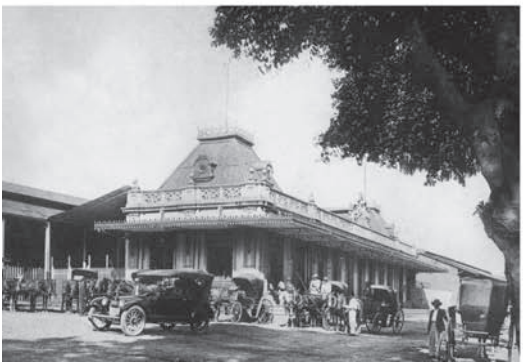 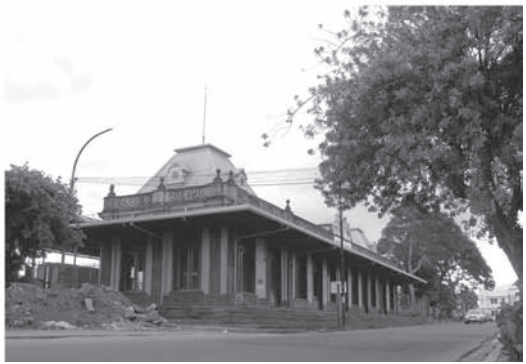

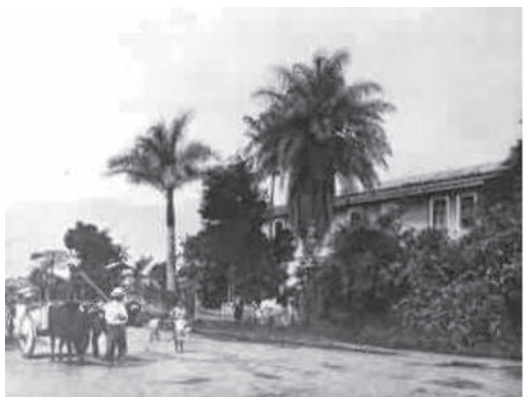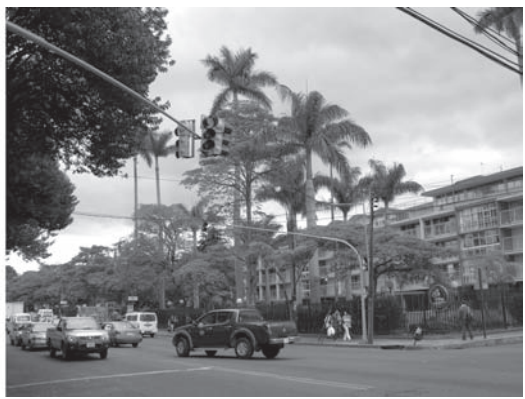

Fig. 1. Images used for the comparison: City of San José, Costa Rica.
NOTE: The original images used in the measurements appear in the Internet edition.

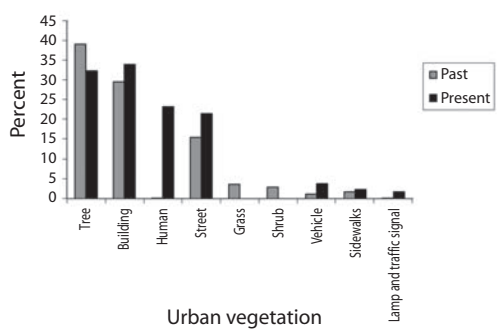

Fig. 2. Change in vegetation, infrastructure, vehicles and humans in photographs from the city of San José, Costa Rica, from a century ago to the present (% cover in photographs).

The families with most species were Asteraceae, Fabaceae, Poaceae, Lamiaceae, Euphorbiaceae, Solanaceae, Cyperaceae, Acanthaceae, Malvaceae, Piperaceae and Verbenaceae (Appendix 1).

## DISCUSSION

Repeat photography was used in Costa Rica by Horn (1989) to assess changes in the paramo habitat (she found very little change) but we were unable to find a similar study of urban vegetation. Nevertheless, studies about plant biodiversity after long periods suggest that our results are not unusual. For example, after 50 years, Brussels has the same total number of plant species that it had in 1940, albeit the individual species change and the same applies to other areas in much longer time spans (Chocholouškováa & Pyšek 2003). We cannot make a comparison of species from *circa* 1900 to *circa* 2000 in San José because urban vegetation has rarely been collected in recent decades.

The increase in human presence in the city landscape is explained by the population growth of downtown San José or *Cantón Central* (from about 30 000 when the first photographs used in this study were taken *circa* 1900 to 356 000 when the sites were rephotographed in 2009; see: Centro Centroamericano de Población and Instituto Nacional de Estadística y Censos 2002). The small increase in the number of vehicles is an underestimation: the 2008 photographs were purposefully taken in low traffic periods to obtain a better view of the scenes.

Central European cities have a mean of 646 plant species/city and larger cities have more species (Pysek 1998), thus, considering that San José was and is a small city, the total of 458 species recorded is within the expected range.

Successful urban plants tend to belong to species adapted to natural habitats with strong sunlight, abundant nitrogen and low water levels (Shaltout & El-Sheikh 2002, Chocholouškováa & Pyšek 2003, Lundholm & Marlin 2006), so the presence of many Asteraceae, Fabaceae, Poaceae, Lamiaceae, Euphorbiaceae, Solanaceae, Cyperaceae, Acanthaceae, Malvaceae, Piperaceae and Verbenaceae is not surprising. Many of the plant species in Costa Rica are introduced (Chacón & Saborío 2005) and here again the situation is similar to that in other countries.

Urban trees mitigate global warming (Abdollahi *et al*. 2000, McPherson *et al*. 2008), significantly reduce urban heat islands (Huang *et al*. 2009) and can sequester about 100kg of air pollutants per hectare of urban forest (Vilela-Lozano 2004). Furthermore, urban vegetation protects many species in the five kingdoms (Dana *et al*. 2002, Smith *et al*. 2006), including valuable rare species (Maurer *et al*. 2000, Williams *et al*. 2009). For these and other reasons, the study and management of urban vegetation is of great importance.

Genetic diversity is low in urban plants and they are less prepared to cope with environmental change (Knapp *et al*. 2009), so periodic monitoring is needed to conserve original species as well as any others in need of protection (Godefroid 2001). Citizens can learn to effectively take advantage of urban vegetation (Garzón *et al*. 2004) and to recognize historical changes in the city scene (e.g. the repeat photography groups in www.flickr.com), not only for cultural reasons, but also to influence the administration of the urban flora by local governments.

Future studies of urban vegetation in San José could investigate these hypotheses:
- Floristic composition results from the interaction of human density, water availability, temperature, altitude and soil (Dana *et al*. 2002, Chocholouškováa & Pyšek 2003, Fanelli & Tescarollo 2006, Altobelli *et al*. 2007).
- Ecological succession starts with ruderal annual plants, followed by perennials (Prach *et al*. 2001).
- Alien species benefit more from human activity (Niggermann 2009).
- With global warming, species adapted to colder climate will become less common and vice versa.

ACKNOWLEDGMENTS

We thank Hubert Blanco (Archivos Nacionales), Colegio Superior de Señoritas, María José Guerra Araus (Herbario Nacional), Sergio Aguilar, Sergio Quesada, Karla Vega, Andrea Sánchez and María Acuña for their assistance. We specially thank Sally P. Horn (University of Tennessee, Knoxville) for suggestions to improve the manuscript and for advice on repeat photography.

RESUMEN

La vegetación urbana es de vital importancia ya que una proporción importante de la población humana vive en ciudades. Sin embargo, esta vegetación es poco estudiada fuera del centro de Europa y se sabe particularmente poco sobre la flora urbana de las ciudades tropicales de Asia, África y América Latina. Aquí presentamos una estimación de cómo ha cambiado la vegetación en la ciudad de San José, Costa Rica, durante un siglo, con la técnica de la fotografía repetida (sobre la base de una colección de fotografías del siglo XIX y principios del siglo XX hechas por José Fidel Tristán y otros) y con los datos del Herbario Nacional de Costa Rica. Encontramos pocos cambios en el paisaje de San José durante el siglo XX. En la ciudad se recolectaron 95 familias y 458 especies entre finales del siglo XIX y principios del XX. Las familias con más especies fueron Asteraceae, Fabaceae, Poaceae, Lamiaceae, Euphorbiaceae, Solanaceae, Cyperaceae, Acanthaceae, Malvaceae, Piperaceae y Verbenaceae. Los resultados son similares a los de Europa, donde el número de especies de plantas a menudo es estable durante largos períodos, aunque las especies individuales varíen.

**Palabras clave:** Flora urbana, efectos de la urbanización, paisaje de ciudad, comparación fotográfica, lista de especies.

# APPENDIX 1

## Species collected in the late 19th and early 20th century

| Species | Date | Collector |
|---|---|---|
| *Acalypha macularis* | 1906-10-28 | Otón Jiménez 46 |
| *Acalypha amentacea* | 1906-12-03 | Otón Jiménez 95 |
| *Acanthus mollis* | 1931-03-25 | Juvenal Valerio 73 |
| *Achimenes grandiflora* | 1888-07-31 | Paul Biolley |
| *Achimenes longiflora* | 1888-07-31 | Paul Biolley |
| *Acmella oppositifolia* | 1889-06-20 | H. Pittier 1091 |
| *Acmella radicans* | 1892-11-30 | A. Tonduz 7185 |
| *Acnistus arborescens* | 1893-06-30 | A. Tonduz |
| *Adiantum colpodes* | 1896-07-31 | A. Tonduz 8064 |
| *Adiantum patens* | 1888-01-15 | H. Pittier 48 |
| *Adiantum concinnum* | 1945-08-06 | Richard W. Holm 991 |
| *Aeschynomene villosa* | 1890-12-31 | Paul Biolley |
| *Agave wercklei* | 1907-02-13 | Anastasio Alfaro |
| *Ageratum conyzoides* | 1931-01-03 | Fernando Solís 2 |
| *Ageratum microcarpum* | 1893-01-31 | A. Tonduz 7281 |
| *Albizia adinocephala* | 1890-07-31 | Paul Biolley |
| *Aleurites moluccanus* | 1906-10-30 | Otón Jiménez 127 |
| *Alternanthera laguroides* | 1896-06-30 | A. Tonduz 10135 |
| *Amaranthus hybridus* | 1888-07-31 | H. Pittier 398 |
| *Amphilophium paniculatum* | 1888-01-31 | H. Pittier 975 |
| *Anagallis pumila* | 1893-02-28 | A. Tonduz |
| *Anemia hirsuta* | 1888-07-31 | Paul Biolley 912 |
| *Anoda cristata* | 1931-01-03 | Fernando Solís 9 |
| *Anredera cordifolia* | 1937-10-21 | José Antonio Echeverría 4007 |
| *Anthurium acutangulum* | 1896-09-30 | A. Tonduz 10360 |
| *Aphelandra deppeana* | 1896-12-31 | A. Tonduz 10401 |
| *Archibaccharis torquis* | 1892-11-30 | A. Tonduz 1535 |
| *Archibaccharis schiedeana* | 1889-12-12 | A. Tonduz 1496 |
| *Arenaria lanuginosa* | 1889-06-20 | A. Tonduz |
| *Arracacia xanthorrhiza* | 1904-11-29 | A. Tonduz 17454 |
| *Arthrostemma ciliatum* | 1888-01-15 | H. Pittier |
| *Asclepias curassavica* | 1892-07-31 | A. Tonduz 436 |
| *Asplenium pumilum* | 1888-07-31 | Paul Biolley 906 |
| *Asplenium aethiopicum* | 1889-06-20 | A. Tonduz 1095 |
| *Asplenium myriophyllum* | 1888-07-31 | Paul Biolley 516 |
| *Asplenium formosum* | 1888-07-31 | Paul Biolley 914 |
| *Aster spinosus* | 1888-07-31 | Paul Biolley 990 |
| *Athyrium palmense* | 1888-07-31 | H. Pittier 905 |
| *Axonopus compressus* | 1890-08-31 | A. Tonduz 2828 |
| *Baccharis braunii* | 1896-09-30 | A. Tonduz 10882 |
| *Begonia cuculata* | 1896-12-31 | A. Tonduz 10399 |
| *Bidens pilosa* | 1888-07-31 | Paul Biolley 980 |
| *Blechnum occidentale* | 1888-07-31 | Paul Biolley 921 |

Species collected in the late 19th and early 20th century

| Species | Date | Collector |
|---|---|---|
| *Blechum brownei* | 1890-12-07 | A. Tonduz 3204 |
| *Blechum pyramidatum* | 1931-03-25 | Juvenal Valerio 79 |
| *Bletia purpurea* | 1903-05-31 | H. Pittier 16726 |
| *Bouvardia glabra* | 1896-09-30 | A. Tonduz |
| *Brassica campestris* | 1889-10-31 | A. Tonduz 3535 |
| *Browallia americana* | 1892-11-30 | A. Tonduz 7183 |
| *Brugmansia candida* | 1892-07-31 | A. Tonduz 442 |
| *Byttneria aculeata* | 1888-07-31 | Paul Biolley |
| *Byttneria carthaginensis* | 1888-09-30 | Paul Biolley |
| *Calea axillaris* | 1889-11-23 | A. Tonduz 1430 |
| *Calea urticifolia* | 1931-01-03 | Fernando Solís |
| *Calliandra calothyrsus* | 1888-08-31 | H. Pittier |
| *Calyptocarpus wendlandii* | 1890-05-15 | A. Tonduz 2334 |
| *Calyptocarpus wendlandii* | 1931-01-03 | Fernando Solís |
| *Calyptranthes pallens* | 1929-05-21 | Manuel Valerio |
| *Campyloneurum irregulare* | 1888-07-31 | Paul Biolley 922 |
| *Campyloneurum xalapense* | 1902-03-31 | Anastasio Alfaro |
| *Canavalia villosa* | 1889-11-23 | A. Tonduz |
| *Canna indica* | 1906-11-30 | 101 |
| *Capsicum annuum* | 1895-09-30 | Paul Biolley |
| *Cassia tonduzii* | 1896-06-30 | A. Tonduz |
| *Cassia patellaria* | 1888-10-31 | H. Pittier |
| *Cassia pubescens* | 1889-09-30 | A. Tonduz |
| *Casuarina cunninghamiana* | 1906-09-29 | Otón Jiménez 84 |
| *Cedrela tonduzii* | 1903-06-30 | H. Pittier |
| *Centaurium quitense* | 1890-12-07 | A. Tonduz |
| *Centradenia inaequilateralis* | 1888-07-31 | Paul Biolley |
| *Cestrum aurantiacum* | 1888-07-31 | Paul Biolley |
| *Cestrum lanatum* | 1893-01-31 | A. Tonduz |
| *Cestrum macrophyllum* | 1901-02-28 | A. Tonduz |
| *Cestrum warscewiczii* | 1888-01-15 | H. Pittier 47 |
| *Cestrum glanduliferum* | 1929-05-05 | A. M. Brenes 37 |
| *Cestrum tomentosum* | 1941-12-09 | José Antonio Echeverría 230 |
| *Cestrum nocturnum* | 1936-04-29 | Juvenal Valerio |
| *Chamaedorea costaricana* | 1905-03-30 | Paul Biolley |
| *Chamaesyce lasiocarpa* | 1892-02-28 | A. Tonduz |
| *Chenopodium ambrosioides* | 1888-01-31 | H. Pittier |
| *Chiococca alba* | 1890-07-31 | Paul Biolley |
| *Chloris virgata* | 1890-09-15 | A. Tonduz 2937 |
| *Chromolaena odorata* | 1896-08-31 | A. Tonduz 10894 |
| *Chrysophyllum cainito* | 1902-09-15 | H. Pittier |
| *Cinchona pubescens* | 1937-09-07 | A. M. Brenes |
| *Cissampelos pareira* | 1896-06-22 | A. Tonduz |

APPENDIX 1 (Continued)

Species collected in the late 19th and early 20th century

| Species | Date | Collector |
| --- | --- | --- |
| *Citharexylum donnell-smithii* | 1895-01-31 | A. Tonduz 9623 |
| *Cleome costaricensis* | 1889-05-20 | A. Tonduz 1086 |
| *Cleome pilosa* | 1889-11-28 | A. Tonduz 1450 |
| *Clerodendrum thomsonae* | 1928-04-29 | Manuel Valerio 64 |
| *Clibadium surinamense* | 1885-07-31 | Paul Biolley 976 |
| *Cnidoscolus urens* | 1903-05-31 | H. Pittier |
| *Coelogyne speciosa* | 1932-11-11 | A. M. Brenes 366 |
| *Commelina diffusa* | 1890-10-31 | A. Tonduz 3049 |
| *Commelina leiocarpa* | 1931-01-03 | Fernando Solís 45 |
| *Consolida ajacis* | 1940-11-20 | María del Carmen Roviralta 10 |
| *Conyza bonariensis* | 1892-10-31 | A. Tonduz 857 |
| *Conyza schiedeana* | 1892-10-31 | A. Tonduz 856 |
| *Conyza canadensis* | 1931-01-03 | Fernando Solís |
| *Cordia spinescens* | 1896-06-30 | A. Tonduz 8879 |
| *Cordia eriostigma* | 1933-04-15 | Fernando Solís 527 |
| *Crinum erubescens* | 1893-08-31 | A. Tonduz 8214 |
| *Crotalaria rotundifolia* | 1892-06-30 | A. Tonduz |
| *Crotalaria vitellina* | 1892-09-30 | A. Tonduz |
| *Crotalaria cajanifolia* | 1888-03-31 | Paul Biolley |
| *Crotalaria acapulcensis* | 1942-09-30 | José Antonio Echeverría |
| *Croton hoffmannii* | 1896-06-30 | A. Tonduz |
| *Croton draco* | 1892-12-31 | A. Tonduz 7261 |
| *Cuphea carthagenensis* | 1891-06-30 | A. Tonduz |
| *Cuphea wrightii* | 1889-07-27 | A. Tonduz |
| *Cuphea appendiculata* | 1888-07-31 | Paul Biolley |
| *Cyclanthera tonduzii* | 1889-11-28 | A. Tonduz 1449 |
| *Cyclospermum leptophyllum* | 1889-08-12 | A. Tonduz 1287 |
| *Cynoglossum amabile* | 1940-11-13 | María del Carmen Roviralta 8 |
| *Cyperus papyrus* | 1906-09-29 | Otón Jiménez 76 |
| *Cyperus flavescens* | 1892-10-31 | A. Tonduz 1531 |
| *Cyperus niger* | 1890-12-07 | A. Tonduz |
| *Cyperus tenuis* | 1892-07-31 | A. Tonduz 433 |
| *Cyperus involucratus* | 1896-09-30 | A. Tonduz 10888 |
| *Cyperus mutisii* | 1906-11-11 | Otón Jiménez 69 |
| *Cyperus hermaphroditus* | 1906-11-11 | Otón Jiménez 68 |
| *Cystopteris fragilis* | 1888-07-31 | Paul Biolley 414 |
| *Dahlia rosea* | 1888-01-15 | H. Pittier 58 |
| *Dalea cliffortiana* | 1890-07-31 | A. Tonduz |
| *Delilia biflora* | 1896-12-31 | A. Tonduz 10893 |
| *Delilia biflora* | 1931-01-03 | Fernando Solís |
| *Dennstaedtia distenta* | 1922-08-07 | Manuel Valerio |
| *Desmodium affine* | 1888-07-31 | Paul Biolley |
| *Desmodium distortum* | 1902-11-30 | H. Pittier |

Species collected in the late 19th and early 20th century

| Species | Date | Collector |
|---|---|---|
| *Desmodium intortum* | 1892-11-26 | A. Tonduz |
| *Desmodium tortuosum* | 1902-12-31 | H. Pittier |
| *Desmodium maxonii* | 1931-01-03 | Fernando Solís |
| *Dicliptera unguiculata* | 1931-01-03 | Fernando Solís 10 |
| *Dieffenbachia oerstedii* | 1890-05-15 | A. Tonduz 2558 |
| *Dietes grandiflora* | 1931-04-11 | A. M. Brenes 89 |
| *Dorstenia contrajerva* | 1896-07-31 | A. Tonduz |
| *Dracaena fragrans* | 1934-02-27 | Juvenal Valerio 1078 |
| *Drymaria cordata* | 1889-10-31 | A. Tonduz |
| *Drymaria villosa* | 1892-06-30 | A. Tonduz |
| *Drymonia serrulata* | 1896-07-31 | A. Tonduz |
| *Dryopteris opposita* | 1888-07-31 | Paul Biolley 413 |
| *Dryopteris patula* | 1888-07-31 | Paul Biolley 923 |
| *Dryopteris litigiosa* | 1896-09-30 | A. Tonduz 132088 |
| *Duranta erecta* | 1934-12-30 | A. M. Brenes 22868 |
| *Echinochloa crus-galli* | 1888-12-31 | H. Pittier 229 |
| *Echinochloa crus-pavonis* | 1890-09-08 | 3016 |
| *Echinocystis coulteri* | 1889-11-30 | A. Tonduz |
| *Eclipta prostrata* | 1892-10-31 | A. Tonduz 9570 |
| *Eleocharis elegans* | 1906-11-11 | Otón Jiménez 74 |
| *Elephantopus mollis* | 18-11-1931 | Manuel Quirós 371 |
| *Elephantopus scaber* | 1889-11-30 | A. Tonduz |
| *Elephantopus spicatus* | 1889-12-12 | A. Tonduz 1494 |
| *Encyclia livida* | 1890-03-10 | A. Tonduz 2176 |
| *Eragrostis ciliaris* | 1890-09-15 | A. Tonduz 2935 |
| *Eragrostis hypnoides* | 1890-05-15 | 2477 |
| *Eragrostis mexicana* | 1890-09-15 | A. Tonduz 2932 |
| *Erechtites hieraciifolius* | 1888-08-26 | H. Pittier 462 |
| *Erigeron cuneifolium* | 1888-11-30 | H. Pittier 649 |
| *Eryngium carlinae* | 1889-06-30 | A. Tonduz 1124 |
| *Erythrina berteroana* | 1899-12-31 | A. Tonduz |
| *Erythrina berteroana* | 1934-03-14 | Álvaro Sánchez |
| *Erythrina crista-galli* | 1931-04-01 | Fernando Solís |
| *Eucalyptus globulus* | 1889-11-30 | A. Tonduz |
| *Eucalyptus globulus* | 1941-08-20 | José Antonio Echeverría 25 |
| *Euphorbia heterophylla* | 1889-10-31 | A. Tonduz |
| *Euphorbia heterophylla* | 1931-01-03 | Fernando Solís |
| *Euphorbia pulcherrima* | 1931-03-25 | A. M. Brenes |
| *Ficus jimenezii* | 1906-11-29 | A. Tonduz 17536 |
| *Ficus velutina* | 1901-11-30 | H. Pittier |
| *Ficus pertusa* | 1901-11-30 | H. Pittier |
| *Ficus goldmanii* | 1897-10-31 | A. Tonduz |
| *Ficus costaricana* | 1901-11-30 | H. Pittier |

# APPENDIX 1 (Continued)

## Species collected in the late 19th and early 20th century

| Species | Date | Collector |
| --- | --- | --- |
| *Fimbristylis dichotoma* | 1893-12-31 | A. Tonduz 1812 |
| *Galeana pratensis* | 1889-07-27 | A. Tonduz 1237 |
| *Galeana pratensis* | 1931-01-03 | Fernando Solís |
| *Galinsoga quadriradiata* | 1888-07-31 | Paul Biolley 994 |
| *Galinsoga quadriradiata* | 1889-08-12 | A. Tonduz 253 |
| *Gonolobus edulis* | 1888-07-31 | Paul Biolley 938 |
| *Gossypium barbadense* | 1893-05-31 | A. Tonduz |
| *Graptophyllum pictum* | 1931-03-25 | A. M. Brenes 24416 |
| *Grevillea banksii* | 1938-01-11 | Jorge León 1003 |
| *Habenaria clypeata* | 1903-08-31 | H. Pittier 16722 |
| *Hamelia patens* | 1889-07-31 | A. Tonduz |
| *Hauya elegans* | 1930-07-08 | Manuel Valerio 51 |
| *Heliotropium indicum* | 1889-10-31 | A. Tonduz 1146 |
| *Heliotropium procumbens* | 1890-05-31 | A. Tonduz 2430 |
| *Heteranthera reniformis* | 1888-10-31 | H. Pittier 553 |
| *Heterocentron glandulosum* | 1892-11-30 | A. Tonduz 1541 |
| *Hibiscus rosa-sinensis* | 1896-12-31 | A. Tonduz |
| *Hibiscus rosa-sinensis* | 1931-04-23 | Vitalia Sáenz 96 |
| *Holmskioldia sanguinea* | 1941-11-10 | José Antonio Echeverría 92 |
| *Homalocladium platycladum* | 1940-12-07 | José Antonio Echeverría 4015 |
| *Hydrocotyle ranunculoides* | 1889-12-31 | A. Tonduz 1688 |
| *Hymenocallis littoralis* | 1905-06-29 | A. Tonduz |
| *Hypericum uliginosum* | 1888-07-31 | Paul Biolley |
| *Hypoxis decumbens* | 1893-06-30 | A. Tonduz 8028 |
| *Hyptis suaveolens* | 1889-10-31 | A. Tonduz |
| *Hyptis mutabilis* | 1893-01-02 | A. Tonduz |
| *Hyptis capitata* | 1931-01-03 | Fernando Solís |
| *Hyptis pectinata* | 1931-01-03 | Fernando Solís |
| *Hyptis urticoides* | 1931-01-03 | Fernando Solís |
| *Indigofera costaricensis* | 1892-10-31 | A. Tonduz |
| *Indigofera tephrosioides* | 1888-07-31 | Paul Biolley |
| *Inga densiflora* | 1934-05-19 | Collector data missing |
| *Inga vera* | 1888-07-31 | H. Pittier |
| *Inga leptoloba* | 1888-05-31 | H. Pittier |
| *Inga oerstediana* | 1888-05-31 | H. Pittier |
| *Ipomoea alba* | 1932-01-06 | Fernando Solís 405 |
| *Ipomoea aristolochiifolia* | 1933-01-01 | Manuel Quirós 608 |
| *Ipomoea batatas* | 1932-01-06 | Fernando Solís 407 |
| *Ipomoea carnea* | 1931-03-25 | Juvenal Valerio 81 |
| *Ipomoea parasitica* | 1931-01-04 | Fernando Solís 91 |
| *Iresine diffusa* | 1893-01-01 | A. Tonduz 7241 |
| *Jaegeria hirta* | 1888-0-731 | Paul Biolley 940 |
| *Jaltomata repandidentata* | 1928-10-05 | A. M. Brenes |

# APPENDIX 1 (Continued)

## Species collected in the late 19th and early 20th century

| Species | Date | Collector |
|---|---|---|
| *Jasminum revolutum* | 1896-09-30 | A. Tonduz |
| *Jatropha gossypiifolia* | 1906-12-03 | Otón Jiménez 94 |
| *Justicia crenata* | 1887-05-31 | Paul Biolley 1013 |
| *Justicia aurea* | 1940-11-13 | María del Carmen Roviralta 9 |
| *Koellikeria erinoides* | 1888-07-31 | Paul Biolley |
| *Kyllinga odorata* | 1892-07-31 | A. Tonduz 434 |
| *Lagerstroemia indica* | 1891-03-31 | H. Pittier |
| *Lantana urticifolia* | 1889-08-31 | A. Tonduz |
| *Lantana hirta* | 1888-07-31 | A. Tonduz |
| *Laportea aestuans* | 1936-11-14 | Juvenal Valerio 1388 |
| *Lasiacis sorghoidea* | 1893-01-01 | A. Tonduz 7234 |
| *Lasiacis swartziana* | 1888-02-20 | H. Pittier 81 |
| *Lasiacis sorghoidea* | 1931-01-03 | Fernando Solís |
| *Lastreopsis effusa* | 1889-07-27 | A. Tonduz 1128 |
| *Leersia hexandra* | 1889-10-31 | A. Tonduz 1383 |
| *Leonurus japonicus* | 1890-09-30 | A. Tonduz 1399 |
| *Leonurus japonicus* | 1931-01-03 | Fernando Solís 25 |
| *Lepidium bipinnatifidum* | 1892-07-31 | A. Tonduz 437 |
| *Ligustrum vulgare* | 1931-04-14 | Fernando Solís 192 |
| *Limonium sinuatum* | 1931-04-11 | A. M. Brenes 87 |
| *Lindernia diffusa* | 1893-02-28 | A. Tonduz |
| *Lipocarpha micrantha* | 1941-11-06 | José Antonio Echeverría 133 |
| *Lippia myriocephala* | 1892-11-26 | Paul Biolley |
| *Lippia alba* | 1888-07-31 | Paul Biolley |
| *Lippia cardiostegia* | 1888-12-31 | H. Pittier |
| *Lobelia xalapensis* | 1892-10-31 | A. Tonduz |
| *Loeselia glandulosa* | 1931-01-04 | Fernando Solís 80 |
| *Lophospermum erubescens* | 1930-06-10 | Vitalia Sáenz |
| *Ludwigia octovalvis* | 1889-10-31 | A. Tonduz |
| *Ludwigia peruviana* | 1889-10-31 | A. Tonduz |
| *Ludwigia peruviana* | 1888-01-15 | H. Pittier 41 |
| *Ludwigia octovalvis* | 1931-01-03 | Fernando Solís 19 |
| *Malachra radiata* | 1889-11-30 | A. Tonduz |
| *Malachra alceifolia* | 1928-10-05 | A. M. Brenes 19 |
| *Malpighia glabra* | 1890-04-30 | A. Tonduz |
| *Malpighia mexicana* | 1891-08-31 | H. Pittier |
| *Malva parviflora* | 1891-08-31 | H. Pittier |
| *Malvaviscus arboreus* | 1892-12-31 | A. Tonduz |
| *Mangifera indica* | 1888-05-31 | H. Pittier 388 |
| *Manihot glaziovii* | 1890-11-30 | A. Tonduz |
| *Marsypianthes chamaedrys* | 1892-07-31 | A. Tonduz |
| *Marsypianthes chamaedrys* | 1931-01-03 | Fernando Solís |
| *Masdevallia ecaudata* | 1890-11-30 | Paul Biolley 3127 |

Species collected in the late 19th and early 20th century

| Species | Date | Collector |
|---|---|---|
| *Maurandia barclayana* | 1928-04-29 | Manuel Valerio |
| *Mauria heterophylla* | 1894-01-31 | Paul Biolley 8475 |
| *Mazus pumilus* | 1937-04-11 | Jorge León |
| *Mecardonia procumbens* | 1892-07-31 | A. Tonduz 7113 |
| *Melampodium divaricatum* | 1892-07-31 | A. Tonduz 6960 |
| *Melampodium gracile* | 1931-01-03 | Fernando Solís |
| *Melanthera nivea* | 1893-11-30 | A. Tonduz 1558 |
| *Melanthera nivea* | 1931-01-03 | Fernando Solís |
| *Mikania micrantha* | 1931-01-03 | Fernando Solís |
| *Mildella intramarginalis* | 1892-11-26 | A. Tonduz 7220 |
| *Mimosa diplotricha* | 1893-02-28 | A. Tonduz |
| *Mimosa pudica* | 1888-07-31 | Paul Biolley |
| *Mimosa sensitiva* | 1889-09-28 | A. Tonduz 1279 |
| *Mirabilis jalapa* | 1931-01-03 | Fernando Solís 27 |
| *Mitracarpus hirtus* | 1931-01-03 | Fernando Solís 36 |
| *Mollugo verticillata* | 1941-03-06 | José Antonio Echeverría 94 |
| *Monstera adansonii* | 1890-07-31 | Paul Biolley 2846 |
| *Montanoa guatemalensis* | 1893-02-28 | A. Tonduz 7331 |
| *Moringa oleifera* | 1896-07-31 | Paul Biolley |
| *Moritzia lindenii* | 1896-09-30 | A. Tonduz 10886 |
| *Muhlenbergia tenella* | 1890-10-09 | A. Tonduz 3015 |
| *Muhlenbergia tenella* | 1906-10-30 | Otón Jiménez 50 |
| *Myriophyllum aquaticum* | 1941-10-18 | Vitalia Sáenz 52 |
| *Myrtus communis* | 1928-04-29 | Manuel Valerio |
| *Nasturtium mexicanum* | 1891-08-31 | H. Pittier 6906 |
| *Nectandra martinicensis* | 1896-06-22 | A. Tonduz 10104 |
| *Nectandra turbacensis* | 1897-06-24 | H. Pittier |
| *Nephrolepis occidentalis* | 1908-10-06 | Humberto Bertolini |
| *Nephrolepis undulata* | 1922-08-07 | Manuel Valerio |
| *Nicotiana tabacum* | 1905-06-29 | A. Tonduz |
| *Ochroma pyramidale* | 1941-08-20 | José Antonio Echeverría 27 |
| *Oenothera biennis* | 1905-06-29 | A. Tonduz 17490 |
| *Oenothera rosea* | 1936-04-29 | Juvenal Valerio 1119 |
| *Oleandra costaricensis* | 1908-09-04 | Humberto Bertolini |
| *Ophioglossum reticulatum* | 1888-07-31 | Paul Biolley 925 |
| *Oplismenus burmannii* | 1890-12-07 | A. Tonduz 3124 |
| *Oplismenus burmannii* | 1906-11-27 | Otón Jiménez 9 |
| *Ornithocephalus bicornis* | 1904-12-30 | Paul Biolley 17512 |
| *Oxalis latifolia* | 1896-07-31 | A. Tonduz |
| *Pachira aquatica* | 1933-04-15 | Fernando Solís 526 |
| *Pachyrhizus erosus* | 1905-05-30 | A. Tonduz |
| *Paspalum costaricense* | 1890-10-31 | A. Tonduz 3028 |
| *Paspalum saccharoides* | 1891-08-31 | H. Pittier 6907 |

Species collected in the late 19th and early 20th century

| Species | Date | Collector |
| --- | --- | --- |
| *Paspalum orbiculatum* | 1889-06-30 | H. Pittier 1183 |
| *Paspalum distichum* | 1888-07-20 | H. Pittier 306 |
| *Paspalum notatum* | 1906-11-29 | Otón Jiménez 29 |
| *Paspalum conjugatum* | 1892-09-30 | A. Tonduz 758 |
| *Passiflora adenopoda* | 1929-08-30 | José María Orozco 42 |
| *Passiflora apetala* | 1896-02-28 | John Donnell Smith |
| *Passiflora edulis* | 1905-04-29 | A. Tonduz |
| *Paullinia barbadensis* | 1893-01-31 | A. Tonduz |
| *Pecluma plumula* | 1888-07-31 | Paul Biolley 920 |
| *Pedilanthus tithymaloides* | 1933-04-06 | Manuel Quirós |
| *Pellaea ovata* | 1908-10-06 | Humberto Bertolini |
| *Peperomia tetraphylla* | 1891-11-30 | A. Tonduz 3198 |
| *Peperomia lanceolatopeltata* | 1892-10-05 | A. Tonduz 7262 |
| *Peperomia cooperi* | 1888-07-31 | Paul Biolley 524 |
| *Peperomia lignescens* | 1898-09-30 | A. Tonduz |
| *Peperomia galioides* | 1892-11-26 | A. Tonduz |
| *Peperomia angularis* | 1893-01-01 | A. Tonduz |
| *Peperomia deppeana* | 1890-07-31 | A. Tonduz |
| *Phaseolus lunatus* | 1893-01-02 | A. Tonduz |
| *Phaseolus lunatus* | 1888-01-15 | H. Pittier 52 |
| *Phaseolus lunatus* | 1932-12-25 | Fernando Solís |
| *Philadelphus myrtoides* | 1889-12-31 | A. Tonduz 1492 |
| *Phlebodium pseudoaureum* | 1888-07-31 | Paul Biolley 900 |
| *Phthirusa pyrifolia* | 1892-08-31 | A. Tonduz |
| *Phthirusa pyrifolia* | 1904-10-30 | A. Tonduz |
| *Phyla scaberrima* | 1889-07-27 | A. Tonduz |
| *Phyllanthus niruri* | 1888-07-20 | H. Pittier |
| *Piper bredemeyeri* | 1889-06-20 | H. Pittier 1088 |
| *Piper hispidum* | 1896-07-31 | A. Tonduz 10154 |
| *Piper aduncum* | 1896-07-31 | A. Tonduz |
| *Piper umbellatum* | 1892-06-30 | A. Tonduz 693 |
| *Pittiera longipedunculata* | 1890-07-12 | A. Tonduz 3200 |
| *Plantago australis* | 1894-07-31 | A. Tonduz |
| *Plantago major* | 1894-07-31 | A. Tonduz |
| *Plectranthus amboinicus* | 1942-04-29 | Manuel Quirós |
| *Pleopeltis macrocarpa* | 1889-07-27 | A. Tonduz 1229 |
| *Pleopeltis astrolepis* | 1892-07-31 | A. Tonduz 7121 |
| *Pleurothallis listerophora* | 1890-07-31 | Paul Biolley 2986 |
| *Poa annua* | 1888-05-31 | H. Pittier 230 |
| *Polyclathra cucumerina* | 1890-07-12 | A. Tonduz 3200 |
| *Polygala platycarpa* | 1888-05-31 | Paul Biolley |
| *Polygala costaricensis* | 1889-11-28 | A. Tonduz 1450 |
| *Polygala violacea* | 1888-07-26 | Paul Biolley 348 |

Species collected in the late 19th and early 20th century

| Species | Date | Collector |
|---|---|---|
| *Polygonum punctatum* | 1892-07-31 | A. Tonduz |
| *Polypodium triseriale* | 1888-07-31 | Paul Biolley 899 |
| *Polypodium polypodioides* | 1905-10-01 | Otón Jiménez 291 |
| *Polypodium triseriale* | 1888-07-31 | Paul Biolley 417 |
| *Ponthieva racemosa* | 1890-12-31 | Paul Biolley 3240 |
| *Portulaca oleracea* | 1890-03-31 | H. Pittier |
| *Prockia crucis* | 1891-05-31 | A. Tonduz |
| *Pseuderanthemum cuspidatum* | 1888-07-31 | Paul Biolley 942 |
| *Psidium solisii* | 1932-12-25 | Fernando Solís 509 |
| *Psidium cattleianum* | 1909-09-21 | A. Tonduz |
| *Psidium cattleianum* | 1931-06-07 | Manuel Quirós |
| *Psychotria pubescens* | 1896-07-31 | A. Tonduz |
| *Pterolepis pumila* | 1888-12-12 | H. Pittier 10 |
| *Punica granatum* | 1905-04-29 | A. Tonduz 17469 |
| *Pycreus rivularis* | 1906-10-11 | Otón Jiménez 75 |
| *Rhynchelytrum repens* | 1906-11-19 | Otón Jiménez 58 |
| *Rhynchosia longeracemosa* | 1892-11-26 | A. Tonduz |
| *Rhynchospora nervosa* | 1906-11-29 | Otón Jiménez 124 |
| *Richardia scabra* | 1896-07-31 | A. Tonduz |
| *Ricinus communis* | 1889-11-30 | A. Tonduz |
| *Rivina humilis* | 1890-10-31 | A. Tonduz |
| *Robinsonella divergens* | 1889-11-28 | A. Tonduz |
| *Robinsonella lindeniana* | 1931-01-12 | Fernando Solís 110 |
| *Rorippa mexicana* | 1887-12-20 | H. Pittier 29 |
| *Rosa multiflora* | 1892-06-30 | A. Tonduz |
| *Rubus urticifolius* | 1888-07-31 | H. Pittier |
| *Rumex obtusifolius* | 1896-07-31 | A. Tonduz 10123 |
| *Russelia sarmentosa* | 1889-07-27 | A. Tonduz |
| *Rytidostylis carthaginensis* | 1888-07-31 | Paul Biolley |
| *Sabazia urticifolia* | 1896-07-31 | A. Tonduz 10143 |
| *Salvia wagneriana* | 1889-11-30 | A. Tonduz 1476 |
| *Salvia occidentalis* | 1893-01-31 | A. Tonduz |
| *Salvia costaricensis* | 1895-10-31 | A. Tonduz |
| *Salvia lasiocephala* | 1893-01-02 | A. Tonduz |
| *Salvia polystachya* | 1888-02-20 | H. Pittier |
| *Salvia alvajaca* | 1892-07-31 | A. Tonduz 701 |
| *Salvia occidentalis* | 1931-01-03 | Fernando Solís |
| *Salvia officinalis* | 1938-03-30 | Manuel Quirós |
| *Salvia polystachya* | 1931-01-03 | Fernando Solís |
| *Salvia tiliifolia* | 1931-01-03 | Fernando Solís |
| *Salvia wagneriana* | 1940-10-19 | María del Carmen Roviralta |
| *Sambucus mexicana* | 1931-07-07 | Fernando Solís 288 |
| *Sambucus mexicana* | 1931-07-07 | Fernando Solís 288 |

Species collected in the late 19th and early 20th century

| Species | Date | Collector |
|---|---|---|
| *Sansevieria hyacinthoides* | 1941-02-18 | José Antonio Echeverría 4057 |
| *Sapium macrocarpum* | 1893-07-31 | A. Tonduz 8209 |
| *Scleria hirtella* | 1888-11-30 | H. Pittier 648 |
| *Scoparia dulcis* | 1889-07-31 | A. Tonduz |
| *Scoparia dulcis* | 1931-09-04 | Fernando Solís |
| *Scutellaria purpurascens* | 1891-06-30 | A. Tonduz |
| *Sechium tacaco* | 1902-10-31 | K. Werklé 16674 |
| *Selaginella serpens* | 1888-07-31 | Paul Biolley 927 |
| *Selaginella cuspidata* | 1888-07-31 | Paul Biolley 929 |
| *Senecio hoffmannii* | 1888-01-15 | H. Pittier 39 |
| *Senna papillosa* | 1896-07-31 | A. Tonduz |
| *Serjania acuta* | 1897-03-31 | A. Tonduz |
| *Setaria geniculata* | 1890-10-04 | A. Tonduz 3008 |
| *Sicyos sertuliferus* | 1932-12-25 | Fernando Solís 1100 |
| *Sida haenkeana* | 1891-05-31 | H. Pittier |
| *Sigesbeckia jorullensis* | 1931-01-03 | Fernando Solís |
| *Simarouba glauca* | 1944-06-28 | José Antonio Echeverría 478 |
| *Smallanthus maculatus* | 1892-07-31 | A. Tonduz 699 |
| *Smilax spinosa* | 1889-11-28 | A. Tonduz 1466 |
| *Solanum lanceolatum* | 1892-09-30 | A. Tonduz |
| *Solanum americanum* | 1890-03-25 | A. Tonduz |
| *Solanum americanum* | 1931-01-03 | Fernando Solís 33 |
| *Solanum umbellatum* | 1934-03-17 | Estrella U. de Pacheco |
| *Solanum wendlandii* | 1932-03-23 | Collector data missing |
| *Sonchus oleraceus* | 1890-11-30 | A. Tonduz 3069 |
| *Sorghum bicolor* | 1888-07-31 | H. Pittier 383 |
| *Spananthe paniculata* | 1896-09-30 | A. Tonduz 10883 |
| *Spananthe paniculata* | 1931-01-03 | Fernando Solís 64 |
| *Spathodea campanulata* | 1901-11-30 | H. Pittier 16212 |
| *Spermacoce assurgens* | 1896-07-31 | A. Tonduz |
| *Spigelia splendens* | 1888-07-31 | Paul Biolley |
| *Sporobolus indicus* | 1890-05-15 | A. Tonduz 2335 |
| *Stachys costaricensis* | 1887-12-20 | H. Pittier 34 |
| *Stachys costaricensis* | 1896-07-31 | A. Tonduz |
| *Stelis tricuspis* | 1888-07-31 | Paul Biolley 949 |
| *Stellaria prostrata* | 1890-10-31 | A. Tonduz |
| *Stellaria ovata* | 1931-01-03 | Fernando Solís 34 |
| *Stemodia verticillata* | 1895-01-31 | A. Tonduz 9621 |
| *Struthanthus orbicularis* | 1893-01-31 | A. Tonduz |
| *Stylosanthes guyanensis* | 1896-11-30 | A. Tonduz |
| *Synadenium grantii* | 1941-11-10 | José Antonio Echeverría 103 |
| *Syzygium jambos* | 1889-10-31 | A. Tonduz |
| *Tabebuia rosea* | 1890-02-28 | Paul Biolley 2215 |

# APPENDIX 1 (Continued)

## Species collected in the late 19th and early 20th century

| Species | Date | Collector |
| --- | --- | --- |
| *Tagetes microglossa* | 1889-11-28 | A. Tonduz 1451 |
| *Talinum paniculatum* | 1889-07-27 | A. Tonduz |
| *Tecoma stans* | 1889-12-12 | A. Tonduz 1495 |
| *Tetrapterys schiedeana* | 1898-07-31 | Paul Biolley |
| *Thalictrum lankesteri* | 1888-07-31 | Paul Biolley |
| *Thelypteris resinifera* | 1886-01-31 | Paul Biolley 1067 |
| *Thunbergia alata* | 1936-04-29 | Juvenal Valerio 1136 |
| *Thunbergia erecta* | 1940-11-13 | María del Carmen Roviralta 5 |
| *Tibouchina longifolia* | 1893-12-31 | A. Tonduz |
| *Tinantia erecta* | 1888-07-31 | H. Pittier 391 |
| *Tithonia longiradiata* | 1890-12-03 | A. Tonduz 3136 |
| *Tournefortia glabra* | 1889-06-20 | A. Tonduz 1089 |
| *Tragia volubilis* | 1892-07-31 | A. Tonduz |
| *Trichilia havanensis* | 1941-01-01 | José Antonio Echeverría 4023 |
| *Trichilia martiana* | 1896-06-22 | A. Tonduz |
| *Trifolium amabile* | 1889-08-12 | A. Tonduz |
| *Tripogandra purpurascens* | 1889-07-27 | A. Tonduz 1252 |
| *Tripogandra serrulata* | 1889-06-20 | A. Tonduz 1186 |
| *Tripsacum andersonii* | 1901-08-31 | Brade 16174 |
| *Verbena litoralis* | 1936-08-19 | Rafael Roig 9 |
| *Vigna adenantha* | 1931-01-03 | Fernando Solís |
| *Vismia baccifera* | 1896-06-30 | A. Tonduz |
| *Xylosma flexuosa* | 1895-01-31 | A. Tonduz |
| *Xylosma velutina* | 1902-03-31 | Paul Biolley |
| *Zephyranthes carinata* | 1893-06-30 | A. Tonduz 8045 |
| *Zexmenia longipes* | 1890-12-03 | A. Tonduz 3135 |
| *Zexmenia costaricensis* | 1892-07-31 | A. Tonduz 7122 |
| *Zexmenia frutescens* | 1909-12-30 | A. Tonduz 17977 |